\documentclass[doublecol]{epl2} 
\usepackage{amsmath}
\usepackage{amsfonts}
\usepackage{psfrag}
\newcommand{\re}{\mathrm{Re}}
\newcommand{\im}{\mathrm{Im}}

\newcommand{\zr}[1]{\mathrm{Re}(Z_{#1})}

\title{How adding zero to the complex relation between production and scattering amplitudes found by van Beveren and Rupp gives the expected real relation}
\shorttitle{Relation between production and scattering in two-body channels}

\author{M. R. Pennington\inst{1} \and D. J. Wilson\inst{1}}
\shortauthor{M. R. Pennington \etal}

\institute{                    
  \inst{1} Institute for Particle Physics Phenomenology - Durham University, 
Durham, DH1 3LE, U.K.
}
\pacs{11.80.Gw}{Multichannel scattering}
\pacs{11.55.Ds}{Exact S-matrices}
\pacs{13.75.Lb}{Meson-meson interactions}

\abstract{If a hadronic production process is dominated by two body final state
interactions, the amplitude $A$ for the production process can be expanded as a sum of the scattering amplitudes $T$ for the relevant two body channels. Van Beveren and Rupp have claimed  
that the unitarity relation ${\rm {Im}} A= T^\dag A$ can be satisfied if the coefficients in this expansion are complex.  We demonstrate that the coefficients have to be real if the scattering amplitudes $T$ satisfy unitarity. Van Beveren and Rupp have merely written real coefficients as a sum of complex numbers.}

\begin{document}

\maketitle

A hadronic production reaction that is dominated by two body final state interactions inevitably has a right hand cut structure that is related to that of the corresponding two body scattering process. Particular examples are the electromagnetic reactions $e^+e^- \to \pi^+\pi^-$ and $\gamma\gamma\to\pi\pi$ with definite angular momentum and isospin.
In a very recent paper, van Beveren and Rupp~\cite{bevrupp} have claimed that the relation between the amplitude $A_k$ for the production process with two body final state $k$ in a given partial wave and the corresponding two body scattering amplitudes with the same final states can be written as
\begin{eqnarray}
A_k\,=\,{\rm Re} (Z_k)\,+\,i\, \sum_{\ell} Z_{\ell}\, T_{k\ell}\quad .
\label{eq_1a}
\end{eqnarray}
where $T_{k\ell}$ is the scattering amplitude for channel $k \to \ell$ in the state with same quantum numbers. 
All the complex functions $A_k$, $Z_k$ and $T_{k\ell}$ depend on the c.m. energy. Eq.~(1) satisfies the unitarity relation 
\begin{eqnarray}
{\rm Im}\,A \;=\; T^{\dag}\, A\quad,
\label{eq_1b}
\end{eqnarray}
 where when we consider an $n$-channel final state $T$ is an $n \times n$ matrix
and $A$ an $n$-component column vector.

In the treatment of van Beveren and Rupp~\cite{bevrupp} the coefficient, or coupling, functions $Z_k$ are complex. They claim eq.~(1) is a new result and that this differs from the conventional wisdom that the production amplitude is related to the relevant scattering amplitudes through wholly real coupling functions, or in the above notation purely imaginary functions $Z_k$. They regard ${\rm Re}(Z_k) = 0$ as a special case. Here we will illustrate that their result is just a complicated way of writing zero as a sum of complex numbers and that with no assumptions beyond 2-body unitarity, the coefficients ${iZ_k}$ can be rewritten in terms of wholly real functions $iZ'_k$, not as a special case, but quite generally{\footnote{The introduction of the factor of \lq i' is just to make the comparison with Ref.~1 easier, and so not complicate the issue further.}.

To make the argument transparent, we consider the case with two final state channels 1 and 2. Then the van Beveren and Rupp result, eq.~(\ref{eq_1a}), for $A_1$ is 
\begin{eqnarray}
 A_1 &=& \zr{1} + iZ_1\, T_{11} + iZ_2\, T_{12}\label{eq_1}\; .\label{eq_3}
\end{eqnarray}
That for $A_2$ is obtained by trivially interchanging the labels 1 and 2.
We now use two simple identities
\begin{eqnarray}
0&=&\frac{T_{11}T_{22}-T_{12}^{\, 2}}{T_{11}T_{22}-T_{12}^{\, 2}}\,-\,1\label{eq_0a}\\
0&=& \frac{T_{11}T_{12}-T_{11}T_{12}}{T_{11}T_{22}-T_{12}^{\, 2}}\; .\label{eq_0b}
\end{eqnarray}
We multiply the right hand side of eq.~(\ref{eq_0a}) by $\zr{1}$
and the right hand side of eq.~(\ref{eq_0b}) by $\zr{2}$ and add these to eq.~(\ref{eq_3}) to give:
\begin{eqnarray}
A_1  &=& \zr{1}\left(\frac{T_{11}T_{22}-T_{12}^{\, 2}}{T_{11}T_{22}-T_{12}^{\, 2}}\right)\nonumber\\
     &+& \zr{2}\left(\frac{T_{11}T_{12}-T_{11}T_{12}}{T_{11}T_{22}-T_{12}^{\, 2}}\right)\nonumber\\
     &+& iZ_1 T_{11} + iZ_2 T_{12} \label{eq_2}
\end{eqnarray}
 Rearranging the first two terms in eq. (\ref{eq_2}),
\begin{eqnarray}
 A_1 &=& \left(\frac{\zr{1} \,T_{22} - \zr{2}\,T_{12}}{T_{11}T_{22} - T_{12}^{\, 2}}\right)T_{11} \nonumber\\
     &+& \left(\frac{\zr{2}\, T_{11} - \zr{1}\,T_{12}}{T_{11}T_{22} - T_{12}^{\, 2}}\right)T_{12} \nonumber\\
\nonumber\\
     &+&  iZ_1 T_{11} + iZ_2 T_{12}\quad.
\end{eqnarray}
This leads to,
\begin{eqnarray}
 A_1 &=& \left[iZ_1 + \left(\frac{\zr{1}\, T_{22} - \zr{2}\,T_{12}}{T_{11}T_{22} - T_{12}^{\, 2}}\right)\right]\,T_{11} \nonumber\\
     &+&\left[iZ_2 + \left(\frac{\zr{2}\, T_{11} - \zr{1}\,T_{12}}{T_{11}T_{22} - T_{12}^{\, 2}}\right)\right]\,T_{12} \,.
\end{eqnarray}
We can clearly write this as
\begin{equation}
A_1 = i\,Z'_1\, T_{11} + i\,Z'_2\, T_{12}
\end{equation}
where the $Z'_k$'s are possibly complex coefficients. We read off
\begin{eqnarray}
 Z'_1&=&Z_1 - i\, \left(\frac{\zr{1}\, T_{22} - \zr{2}\, T_{12}}{T_{11}T_{22} - T_{12}^{\, 2}}\right)\\
%\end{equation}
%\begin{equation}
 Z'_2&=&Z_2 - i\, \left(\frac{\zr{2}\, T_{11} - \zr{1}\, T_{12}}{T_{11}T_{22} - T_{12}^{\, 2}}\right) .
\end{eqnarray}
We now consider the real part of the $Z'_k$, for instance with $k=1$
\begin{eqnarray}
\re(Z'_1) &=& \zr{1} + \zr{1}\,\im\left(\frac{T_{22}}{T_{11}T_{22}-T_{12}^{\, 2}}\right)\nonumber\\
          &-& \zr{2}\,\im\left(\frac{T_{12}}{T_{11}T_{22}-T_{12}^{\, 2}}\right)\quad .
\end{eqnarray}
Applying the unitarity condition $\im\mathbf{T}^{-1}=-\mathbb{I}$, we have:
\begin{eqnarray}
\im [T^{-1}]_{ij}&=&0 \quad {\rm with}\quad i\ne j \\
\im [T^{-1}]_{jj}&=&-1
\end{eqnarray}
hence $\re (Z'_1) = 0$, therefore $Z'_1$ is purely imaginary, and a similar argument follows for $Z'_2$.
Consequently, two-body unitarity requires that the van Beveren and Rupp equation with seemingly complex coefficients $Z_k$ can be written in terms of purely imaginary functions $Z'_k$. Thus, as was long ago recognised~\cite{watson,aitch,amp,cahn,chung}, the coupling functions $iZ'_k$ are wholly real.
\begin{figure}%[h]
 \onefigure[width=8.cm]{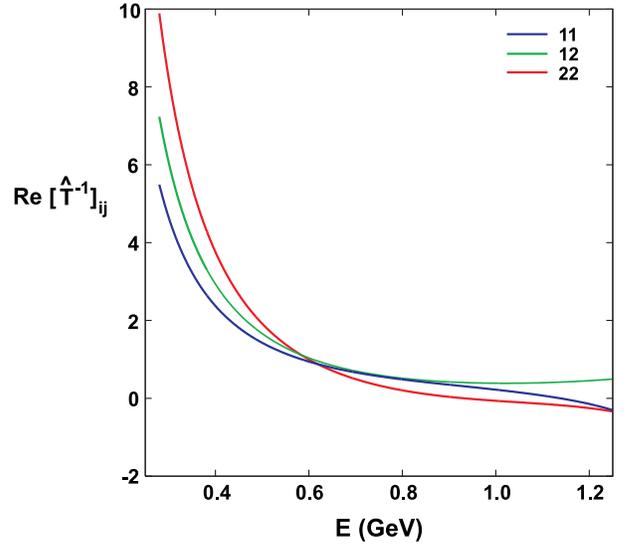}
 \caption{Illustration of the Real parts of the functions $\left[\hat{T}^{-1}\right]_{k\ell}$ that enter the ${\rm Im} Z'_k$ of eqs.~(10,11) for typical hadronic channels, where $1\,=\,\pi\pi$ and $2\,=\,{\overline K}K$ from Ref.~\cite{boglione}. As expected these coefficient functions are perfectly smooth, having no right hand cut singularities. For convenience, the Adler zeroes in each meson-meson scattering amplitude have been divided out. This is the definition of $\hat{T}$.}
\end{figure}

It is important to note that the factors $\left[T^{-1}\right]_{k\ell}$ ( for instance, $\left[T^{-1}\right]_{11} = T_{22}/(T_{11} T_{22} - T_{12}^{\, 2})$), which
relate the coefficients $iZ_k$ to the real functions $iZ'_k$, contain no particle poles. As an illustration we show in fig.~1 these factors for the well studied $I = J = 0$ channel with $\pi\pi$ and ${\overline K}K$ final states for typical model amplitudes~\cite{boglione}, where for convenience we have divided out the Adler zero factors in each of the meson scattering amplitudes.  As expected these fuctions are smooth having no right hand cut singularities, and easily parameterisable as simple polynomials in $E$.

One of the features of writing the production amplitudes as
\begin{eqnarray}
A_k\;=\;i\,\sum_{\ell=1}\;Z'_{\ell}\, T_{\ell k}
\end{eqnarray}
where the functions $iZ'_k$ are wholly real with no right hand cut singularities, is that they pass each threshold in an obviously continuous way. This is not the case for the amplitudes parametrised in terms of the complex functions $Z_k$ that van Beveren and Rupp propose. This is most easily illustrated by considering how the different representations with $n$ channels continue below the 2nd threshold into the region of elastic unitarity.  There $T_{11}\,=\sin\delta e^{i\delta}$ and so 
\begin{eqnarray}
\nonumber A_1&=& {\rm Re}(Z_1)\,+\, i\,Z_1\,T_{11}\;+\;i\,\sum_{\ell=2}\,Z_{\ell}\, T_{\ell 1}\\
\nonumber &=& \left[ {\rm Re}(Z_1) \cos \delta\,+\,{\rm Im}(Z_1) \sin\delta \right]\, e^{i\delta}\;+\;i\,\sum_{\ell=2}\,Z_{\ell}\, T_{\ell 1} .
\end{eqnarray}
Clearly if we have a one channel representation with $Z_{\ell}=0$ for $\ell \ge 2$, Watson's theorem~\cite{watson}, which requires the amplitude $A_1$ has the phase of $T_{11}$ in the region of elastic unitarity, is of course satisfied as the coefficient of the explicit $\exp(i\delta)$ is real. However, Watson's theorem must remain satisfied however many channels we include in the analysis. Typical representations like the $K$-matrix (and its analytic generalisations~\cite{boglione}) continue all the $T_{\ell 1}$ amplitudes in a way that ensures they all have the elastic phase, $\delta$, in the region of elastic unitarity. We see this automatically happens if the $iZ_{\ell}$ are real. But for van Beveren and Rupp~\cite{bevrupp} $iZ_{\ell}$ are complex functions, which they specifically modell as Hankel functions dependent upon the c.m. 3-momentum of the final state. 
With the expected analytic continuation of amplitudes $T_{\ell 1}$, these are all  real below the $\ell$-th threshold. Thus the van Beveren and Rupp functions $Z_{\ell}$ must have right hand cut singularities to be continuable below each threshold and still satisfy unitarity.

Since eq.~(\ref{eq_1a}) is nothing more than a statement that a complex number, which is a two dimensional vector, can be written as the sum of any other two independent complex functions, with real coefficients, then this immediately generalises to the full $n \times n$ scattering matrix. 

Van Beveren and Rupp have literally added zero to the known result and obtained something more complicated: a complication not required by 2-body unitarity.

%
%Insert here the text.
%See fig.~\ref{fig.1}, table~\ref{tab.1} and eq.~(\ref{eq.1}).
%See also~\cite{b.a,b.b}.
%\begin{equation}
%\label{eq.1}
%0\neq1
%\end{equation}
%
% 
% \begin{figure}
% \onefigure{epl-template.eps}
% \caption{Figure caption.}
% \label{fig.1}
% \end{figure}
% 
% 
% \begin{table}
% \caption{Table caption.}
% \label{tab.1}
% \begin{center}
% \begin{tabular}{lcr}
% first  & table & row\\
% second & table & row
% \end{tabular}
% \end{center}
% \end{table}

\acknowledgments
\noindent DJW thanks the U.K.~Science \& Technology Facilities Council (STFC) for a studentship. We acknowledge the partial support of the EU-RTN Programme, Contract No. MRTN--CT-2006-035482, \lq\lq Flavianet''.


\begin{thebibliography}{99}
\bibitem{bevrupp} E.~van~Beveren and G.~Rupp, \lq\lq The complex relation between production and scattering amplitudes'', arXiv:0710.5823 [hep-ph].
\bibitem{watson} K.~M.~Watson, 
%``The Effect Of Final State Interactions On Reaction Cross-Sections,''
  {\it Phys.\ Rev.}  {\bf 88} (1952) 1163
\bibitem{aitch} I.~J.~R.~Aitchison, {\it Nucl. Phys.} {\bf A189} (1972) 417.
\bibitem{amp} K.~L.~Au, D.~Morgan and M.~R.~Pennington, Proc. 21st Rencontre de Moriond on Strong Interactions and Gauge Theories, Les Arcs, France (March 1986) (Moriond 1986: Hadronic v.~2): p. 455; {\it Phys. Rev.} {\bf D35} (1987) 1633.
\bibitem{cahn} R.~H.~Cahn and P.~V.~Landshoff,
  %``Mystery Of The Delta (980),''
  {\it Nucl.\ Phys.}   {\bf B266} (1986) 451.
\bibitem{chung} S.~U.~Chung {\it et al.}, {\it Annalen Physik} {\bf 507} (1995) 404.
\bibitem{boglione} M.~E.~Boglione, Proc. 6th Conf. on Quark Confinement and the Hadron Spectrum, Villasimius, Sardinia, Italy, (Sept. 2004) AIP Conf. Proc. {\bf 756} (2005) 318 [hep-ph/0412034];
M.~E.~Boglione and M.~R.~Pennington (in preparation).
\end{thebibliography}
\end{document}